\documentclass[12pt]{article}

\usepackage{scicite}
\usepackage{times}
\usepackage{physabbrv}
\usepackage{graphicx}
\usepackage{amsmath,amssymb}
\usepackage{url}
\usepackage{xcolor}
\usepackage{pdfpages}
\usepackage[colorlinks=false,citecolor=blue,linkcolor=blue]{hyperref}

\newcommand{\nper}{369} 
\newcommand{\nperdiscard}{4} 
\newcommand{\nnonper}{2529} 

\newenvironment{sciabstract}{%
\begin{quote} \bf}
{\end{quote}}

\title{The Sun is less active than other solar-like stars}

\author{
Timo Reinhold,$^{1\ast}$ 
Alexander I.\ Shapiro,$^{1}$ 
Sami K.\ Solanki,$^{1,2}$ \\
Benjamin T.\ Montet,$^{3}$ 
Natalie A.\ Krivova,$^{1}$ 
Robert H.\ Cameron,$^{1}$ \\
Eliana M.\ Amazo-G\'{o}mez$^{1,4}$ \\
\normalsize{$^{1}$Max-Planck-Institut f\"ur Sonnensystemforschung,}\\  
\normalsize{Justus-von-Liebig-Weg 3, 37077 G\"ottingen, Germany}\\
\normalsize{$^{2}$School of Space Research, Kyung Hee University, Yongin, Gyeonggi, 446-701, Korea}\\
\normalsize{$^{3}$ School of Physics, University of New South Wales, Sydney, NSW 2052, Australia} \\
\normalsize{$^{4}$ Georg-August Universit\"{a}t G\"{o}ttingen, Institut f\"{u}r Astrophysik,  37077 G\"{o}ttingen, Germany} \\
\normalsize{$^\ast$E-mail: reinhold@mps.mpg.de}
}

\date{}

\begin{document} 

\baselineskip24pt

\maketitle 

\begin{sciabstract}
Magnetic activity of the Sun and other stars causes their brightness to vary. We investigate how typical the Sun's variability is compared to other solar-like stars, i.e. those with near-solar effective temperatures and rotation periods. By combining four years of photometric observations from the Kepler space telescope with astrometric data from the Gaia spacecraft, we measure photometric variabilities of \nper~solar-like stars. Most of the solar-like stars with well-determined rotation periods show higher variability than the Sun and are therefore considerably more active. These stars appear nearly identical to the Sun, except for their higher variability. Their existence raises the question of whether the Sun can also experience epochs of such high variability.
\end{sciabstract}

Stars like the Sun have a magnetic field in their interiors, driven by a self-sustaining dynamo process \cite{Reiners2012_LR}. When the magnetic field becomes unstable it can emerge from the stellar surface, leading to the appearance of magnetic features, such as bright faculae and dark starspots. As stars rotate, the transits of these magnetic features across their visible surface, and the temporal evolution of these features, lead to stellar brightness variations. Such variations have been extensively studied for the Sun \cite{Solanki2013}, where they have an amplitude of up to 0.3\% of the sunlight integrated over the entire spectrum, i.e., the total solar irradiance (TSI). Solar variability affects Earth's climate on decadal and longer timescales \cite{bondetal2001}, and Earth's atmospheric chemistry on daily and monthly timescales \cite{Wang2015}. Sufficiently precise solar brightness measurements have only been available since the advent of dedicated spaceborne missions in 1978 \cite{koppetal2014}. Records of sunspot areas and positions can be used to reconstruct brightness variations back to 1878 \cite{SAT_T2_1}. Sunspot counts, the longest record of regular observations of solar magnetic activity, extend back to the onset of telescopic observations around the year 1610 \cite{Activity_LR}. Solar activity can be reconstructed over longer periods, up to 9000~years, from cosmogenic isotopes \cite{Wu2018}.

We take an alternative approach, by comparing the Sun's activity to other solar-like stars \cite{Basri2013, Radick2018}. 
Stellar magnetic activity and photometric variability are strongly correlated (e.g., \cite{Karoff2016}). The same applies to the Sun, for which there is a close correlation between proxies for solar magnetic activity and photometric variability \cite{Salabert2017,Supp}. 
There is an ongoing debate whether solar photometric variability is smaller than the variability of other stars with near-solar effective temperatures and a similar level of magnetic activity \cite{Foukal1994,Radick2018,Witzke2018}. With the advent of planet-hunting missions, in particular the Kepler space telescope \cite{Borucki2010}, this topic enjoyed a new lease on life. For example, the Sun has been found to be photometrically quieter than most of the stars observed by the Kepler space telescope \cite{Gilliland2011}. In contrast, the TSI has a similar level of variability compared to a sample of main-sequence stars with near-solar (and lower) effective temperatures in the Kepler field \cite{Basri2013}. Those studies could not constrain their samples to near-solar rotation periods, due to a lack of available measurements. This may have affected their results,
because the stellar rotation period and effective temperature are related to the action of the dynamo, and therefore the level of magnetic activity \cite{Reiners2012_LR}.

To compare solar photometric variability with other stars, we focus on Kepler observations of main-sequence stars with near-solar fundamental parameters and rotation periods. The stellar fundamental parameters we consider are the effective temperature $T_\text{eff}$, surface gravity $\log\,g$, and metallicity $[Fe/H]$ \cite{Huber2014,Mathur2017}. We adopt a parameter catalog \cite{Mathur2017} that is based on Kepler data release~25 (DR25). Rotation period measurements are available for thousands of stars observed during the Kepler mission \cite{Reinhold2013,McQuillan2014}. 
We adopt a catalog of 34,030 stars with determined rotational periods, and 99,000 stars for which no rotation periods were detected [\cite{McQuillan2014}, their tables 1 and 2]. We refer to these as the "periodic" and the "non-periodic" samples. From both samples we select stars with effective temperatures in the range 5500--6000\,K (the value for the Sun (subscript $\odot$) is $T_{\text{eff},\odot}=5780$\,K) and surface gravities $\log\,g>4.2$ (Sun: $\log\,g_{\odot}=4.44$) to focus on solar-like main-sequence stars. The surface gravity cut removes evolved stars, which are inactive, so may have diluted the variability of solar-like stars found in previous analyses \cite{McQuillan2014}. For the periodic sample, we select rotation periods in the range 20--30~days (Sun: $P_{\text{rot},\odot}=24.47$~days sidereal rotation period). 

\begin{figure}
  \centering
  \includegraphics[width=0.7\textwidth]{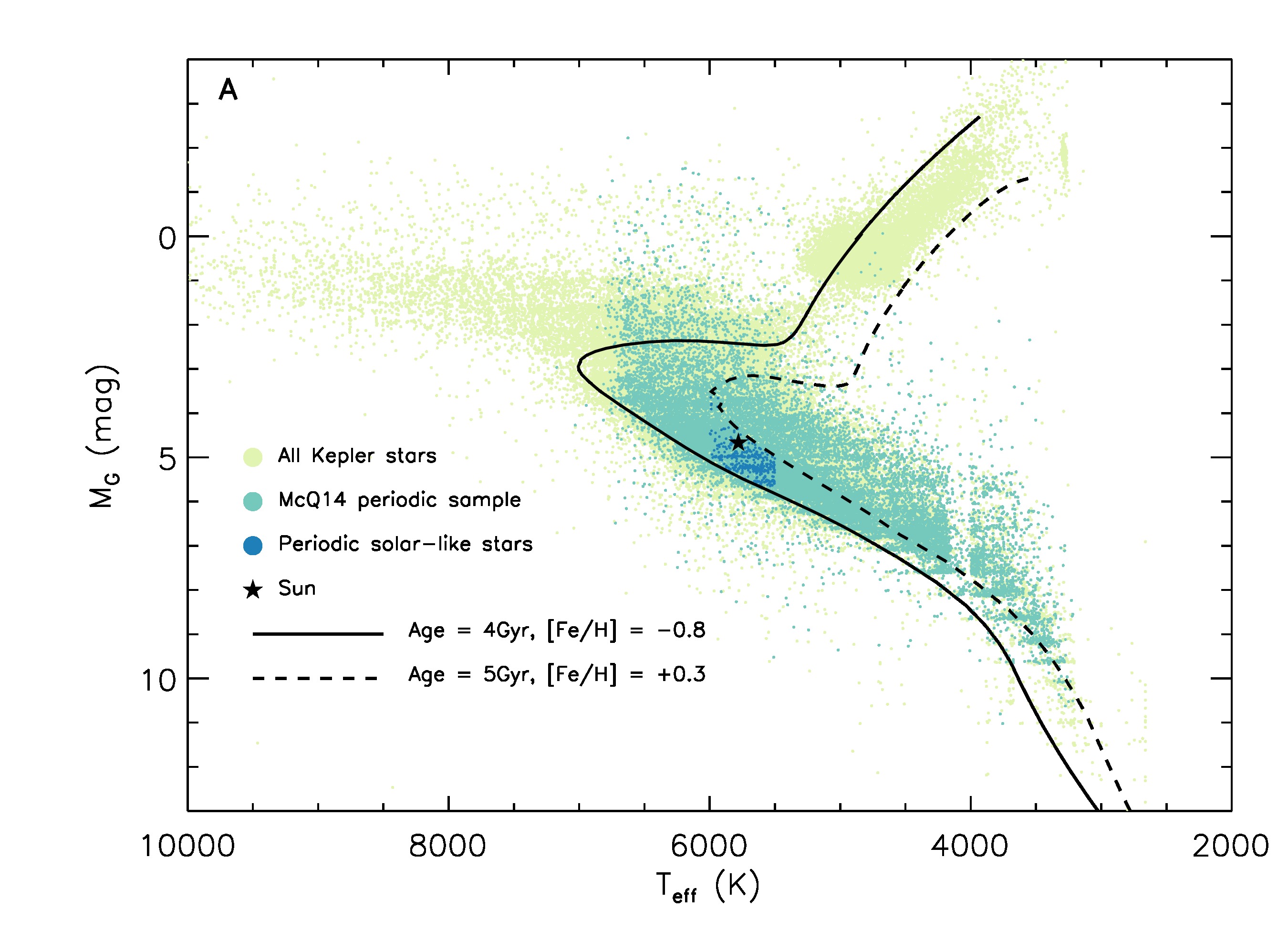}
  \includegraphics[width=0.7\textwidth]{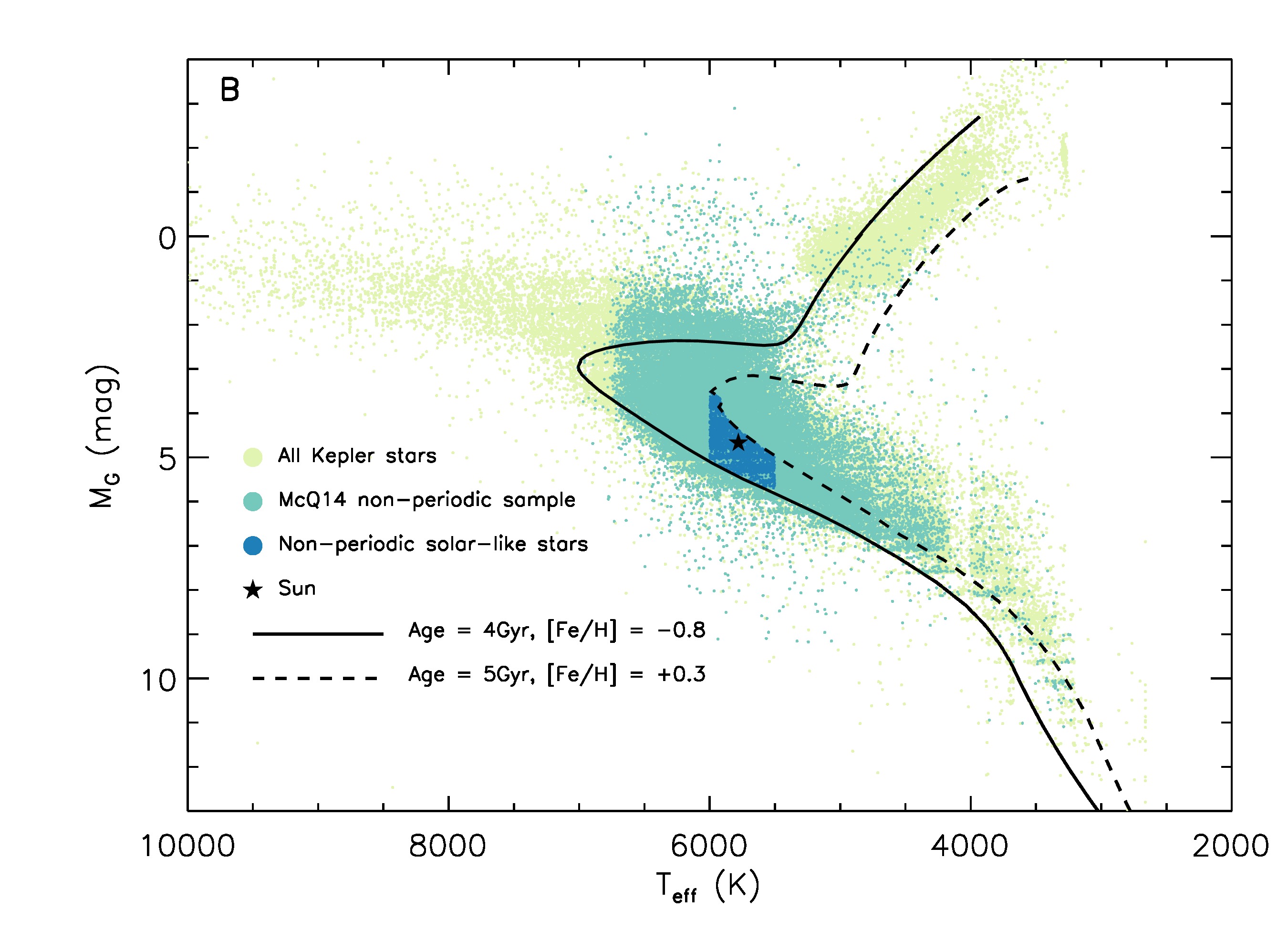}
  \caption{\textbf{Hertzsprung-Russell diagrams of our samples.} The periodic \textbf{(A)} and non-periodic \textbf{(B)} samples \cite{McQuillan2014} (McQ14) are shown in dark green, and the stars that meet our selection criteria are overplotted in blue. The solid black line is a 4\,Gyr isochrone with a metallicity $[Fe/H] = -0.8$, and the dashed black line is a 5\,Gyr isochrone with a metallicity $[Fe/H] = 0.3$. The Sun is indicated by a black star.}
  \label{HRD}
\end{figure}

We further restrict the samples using astrometric data from the Gaia spacecraft \cite{Gaia_mission}. Using the sample stars' apparent magnitudes, distance measurements \cite{Bailer-Jones2018}, and interstellar extinctions from Gaia data release~2 (Gaia~DR2 \cite{Gaia_DR2}), we construct a Hertzsprung-Russell diagram (HRD) by computing the absolute Gaia G-band magnitudes $M_\text{G}$ (Fig.~\ref{HRD}). The absolute magnitudes of our samples are restricted by selecting stars from the HRD with near-solar ages between 4--5\,Gyrs (Sun: 4.57\,Gyr) and metallicities in the range -0.8~dex to 0.3~dex. This is realized by fitting isochrones (i.e. evolutionary tracks of constant age \cite{Supp}) to the HRD, and then selecting periodic and non-periodic stars between a lower isochrone of 4\,Gyr and metallicity of $[Fe/H]=-0.8$, and an upper isochrone of 5\,Gyr and metallicity of $[Fe/H]=0.3$ (Fig.~\ref{HRD}A-B). Stellar variability depends only weakly on metallicity \cite{Supp}, so a stricter metallicity constraint does not affect our results; we therefore use this broad range to improve the statistics. The Sun is slightly more luminous than the majority of selected periodic and non-periodic stars (Fig.~\ref{HRD}), because 79\% of these stars have metallicities lower than the solar value.

We consider stars in our periodic sample to be solar-like, i.e. they have near-solar fundamental parameters and rotation periods. The non-periodic stars are considered only pseudo-solar, because their rotation periods are not known. We then discard stars fainter than 15th magnitude (in the Kepler band) due to their high noise level, which could mask the stellar variability. 
After applying all these selection criteria, our final samples contain \nper~solar-like stars with determined rotation periods, and \nnonper~pseudo-solar stars without a detected period.

To quantify the magnetic activity of the Sun and the selected stars, we compute their photometric variability using the variability range $R_\text{var}$. This quantity is defined as the difference between the 95th and 5th percentile of the sorted flux values (normalized by its median) in a light curve (i.e. the temporal record of the stellar flux) \cite{Basri2010}. Our $R_\text{var}$ values are based on the Kepler Presearch Data Conditioning (PDC) and maximum a priori (MAP) detrended data \cite{Smith2012}. We selected the PDC-MAP data after considering how the different Kepler data products may affect our results \cite{Supp}.

We found that $R_\text{var}$ in the periodic sample shows a weak dependence on effective temperature, rotational period, and metallicity (Fig.~\ref{correction}), even though these were constrained to narrow ranges by our selection criteria. We therefore corrected the $R_\text{var}$ measurements of the periodic stars for these dependencies, and normalized them to the values of the solar fundamental parameters using a multivariate analysis \cite{Supp}. For \nperdiscard~of the \nper~periodic stars, this process returned negative $R_\text{var}$ values, indicating an over-correction. Those \nperdiscard~stars were discarded. For the non-periodic sample, $R_\text{var}$ does not correlate with the fundamental parameters (Fig.~\ref{no_correction}), so no correction was applied.

\begin{figure}
  \centering
  \includegraphics[width=0.7\textwidth]{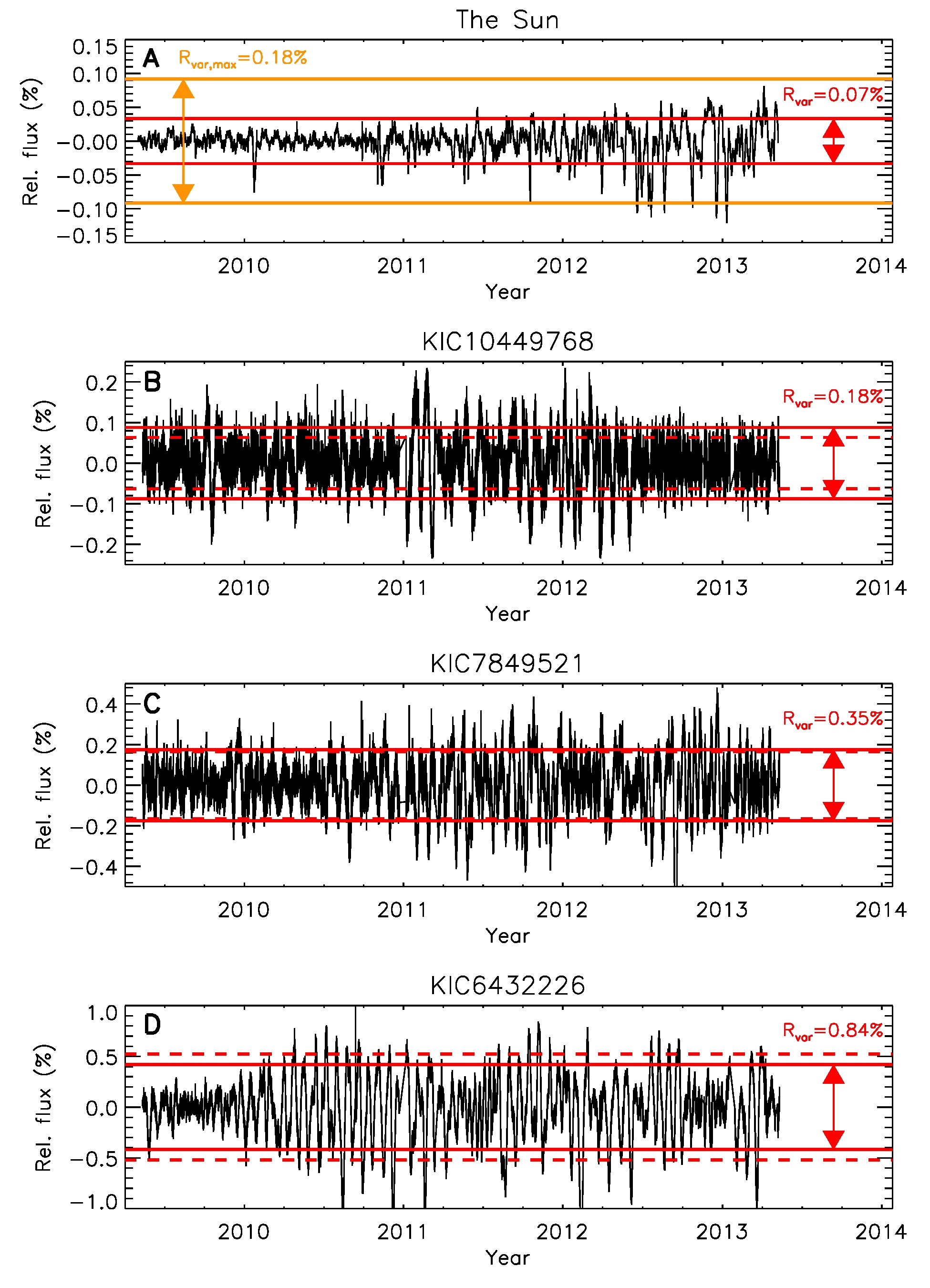}
  \caption{\textbf{Light curves of the Sun (A) and three stars from the periodic sample (B-D).} \textbf{(A)} Solar TSI data taken at the same epoch as the Kepler observations. The TSI data were detrended by cutting the 4-year time series into 90-day segments, dividing by the median flux and subtracting unity. \textbf{(B-D)} Three examples of stars with different variabilities. The variability ranges $R_\text{var}$ are indicated by the differences between the horizontal red lines before (dashed) and after (solid) correction for the variability dependence on the fundamental parameters. The solid orange lines in \textbf{(A)} mark the maximum solar variability range (Fig.~\ref{Rvar_dist} and \cite{Supp}). The panels have different y-scales.}
  \label{lcs}
\end{figure}

Fig.~\ref{lcs} shows three example stellar light curves and solar TSI data \cite{Supp} taken at the same epoch as the Kepler observations. TSI data have been demonstrated to be suitable for the direct comparison with variability observed in the Kepler passband \cite{Basri2013, Supp}. While the star KIC\,10449768 exhibits variability that is similar to the maximum observed solar variability \cite{Supp}, the other three stars in Figure~\ref{lcs} have much higher variability.

\begin{figure}
  \centering
  \includegraphics[width=1.0\textwidth]{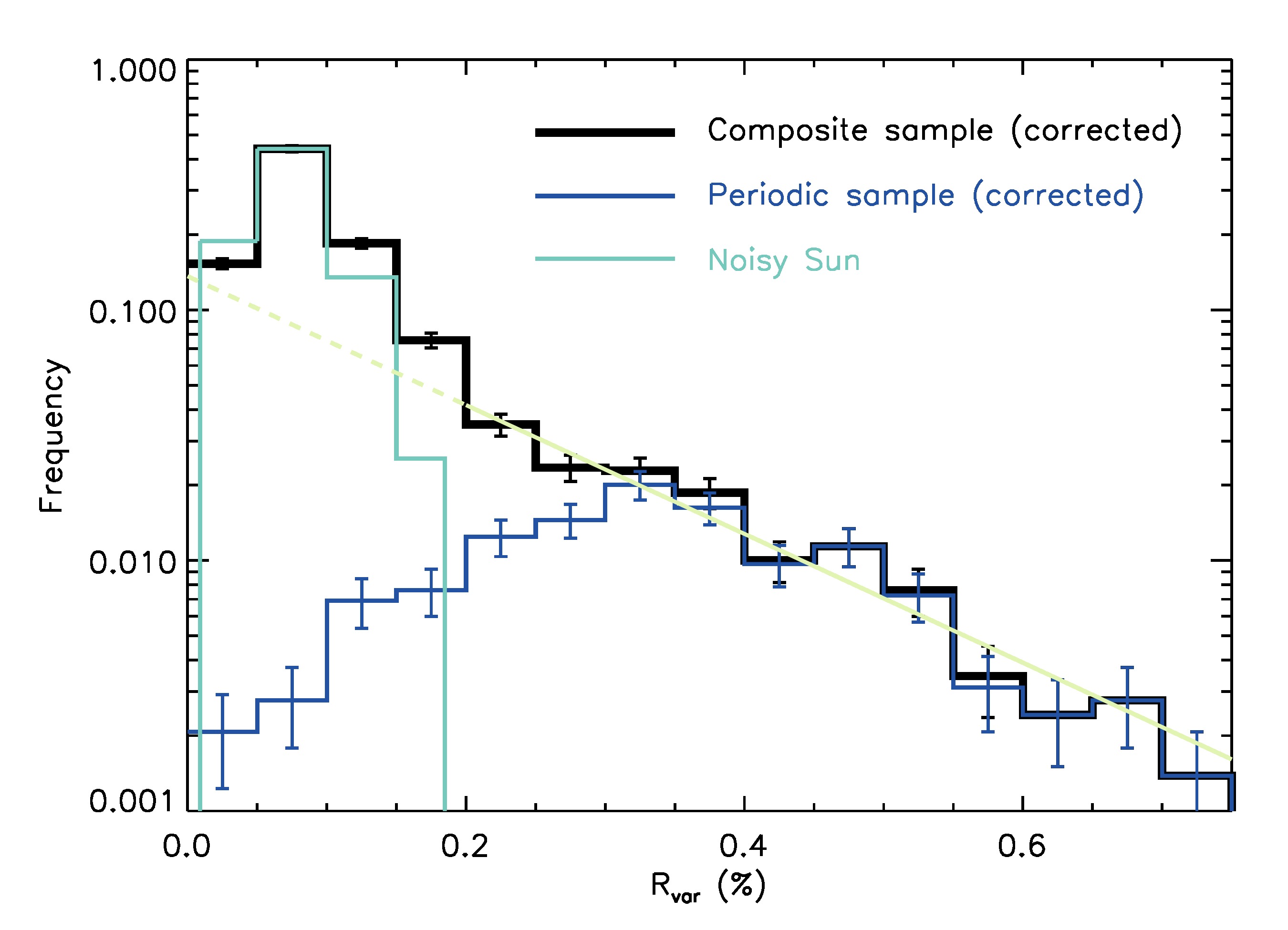}
  \caption{\textbf{Solar and stellar variability distributions on a logarithmic scale.}
  The distributions of the variability range $R_\text{var}$ are plotted for the composite sample (black), the periodic sample (blue), and the Sun over the last 140 years (green). Error bars indicate the statistical uncertainties $\sqrt{N}$ for the number of stars in each bin, $N$, for the composite and the periodic samples. The yellow line shows an exponential model $a_0\,10^{a_1 R_\text{var}}$ fitted to the variability distribution of the (corrected) composite sample ($R_\text{var}>0.2\%$, solid line) and its extrapolation to low variabilities ($R_\text{var}<0.2\%$, dashed line). The solar distribution was normalized to the maximum of the composite sample. The first and last bins of the solar distribution were reduced in width to stop at the minimum and maximum values of solar variability over the last 140 years, respectively.}
  \label{Rvar_dist}
\end{figure}

Figure~\ref{Rvar_dist} shows the distribution of $R_\text{var}$ for the Sun, the periodic stars, and a composite sample of the periodic and non-periodic samples combined. To compare the Sun with the stars observed by Kepler, we simulated how it would have appeared in the Kepler data by adding noise to the TSI time series (Fig.~\ref{Rvar_area}). The variability range was then computed for 10,000 randomly selected 4-year segments from $\sim$140~years of reconstructed TSI data \cite{Supp}.

The activity distribution of the composite sample (Fig.~\ref{Rvar_dist}) does not separate into distributions of periodic and non-periodic stars, but appears to represent a single physical population of stars. Fitting an exponential function $y=a_0\,10^{a_1 R_\text{var}}$ to the variability distribution of the (corrected) composite sample with $R_\text{var}>0.2\%$ yields $a_0=0.14\pm0.02$ and $a_1=-2.27\pm0.17$. The subsample of periodic stars mostly populates the high variability portion of the full distribution in Figure~\ref{Rvar_dist}, whereas the low variability portion mostly contains stars from the non-periodic sample. The solar $R_\text{var}$ distribution is consistent with the majority of low-variability stars, in line with previous studies \cite{Basri2013}. Determining the solar rotation period from photometric observations alone is challenging \cite{Aigrain2015,Sasha_NAT,Timo2019}. Consequently, the Sun would probably belong to the non-periodic sample if it were observed by Kepler, and we find that the level of solar variability is typical for stars with undetected periods (Fig.~\ref{Rvar_dist}). The Sun would appear as a rather normal star of the non-periodic sample if it had been observed with Kepler. However, our composite sample contains stars that might have quite different rotation periods, even though they have near-solar fundamental parameters.

In contrast, the variability of stars in the periodic sample has a different distribution. While there are some periodic stars with variabilities within the observed range covered by the Sun, the variability amplitude for the majority of periodic stars lies well above the solar maximum value of the last 140~years. Consequently, most of the solar-like stars that have measured near-solar rotation periods appear to be  more active than the Sun. The variability of the periodic stars at the solar effective temperature, rotation period, and metallicity is $R_\text{var}=0.36$\% (Fig.~\ref{correction}), which is about 5~times higher than the median solar variability $R_{\text{var},\odot}=0.07$\%, and 1.8~times higher than the maximum solar value $R_{\text{var},\odot} \lesssim 0.20$\%. All these stars have near-solar fundamental parameters and rotational periods, so this implies that their values do not uniquely determine the level of any star's magnetic activity. This result is consistent with the detection of flares with energies several orders of magnitude higher than solar flares (i.e., superflares) on other solar-type stars \cite{Maehara2012,Notsu2019}.


We suggest two interpretations of our result. First, there could be unidentified differences between the periodic stars and non-periodic stars (like the Sun). For example, it has been proposed that the solar dynamo is in transition to a lower activity regime \cite{Metcalfe2016,Metcalfe2017} due to a change in the differential rotation inside the Sun. According to this interpretation, the periodic stars are in the high-activity regime, while the stars without known periods are either also in transition, or are in the low-activity regime. The second possible interpretation is that the composite sample in Fig.~\ref{Rvar_dist} represents the distribution of possible activity values the Sun (and other stars with near solar fundamental parameters and rotational periods) can exhibit. In this case, the measured solar distribution is different only because the Sun did not exhibit its full range of activity over the last 140~years. Solar cosmogenic isotope data indicate that in the last 9000~years the Sun has not been substantially more active than in the last 140~years \cite{Wu2018}. There are several ways for this constraint to be reconciled with such an interpretation. For example, the Sun could alternate between epochs of low and high activity on timescales longer than 9000~years. Our analysis does not allow us to distinguish between these two interpretations.

\newpage
\bibliography{biblothek}
\bibliographystyle{Science}
\newpage
\subsection*{Acknowledgments}
\textbf{Acknowledgments:} We thank the three anonymous referees for constructive criticism and useful advice, which helped to greatly improve the paper. We thank the International Space Science Institute, Bern, for their support of science team~446 and the resulting helpful discussions. This paper includes data collected by the Kepler mission. Funding for the Kepler mission is provided by the NASA Science Mission directorate. This work has made use of data from the European Space Agency (ESA) mission
{\it Gaia} (\url{https://www.cosmos.esa.int/gaia}), processed by the {\it Gaia}
Data Processing and Analysis Consortium (DPAC,
\url{https://www.cosmos.esa.int/web/gaia/dpac/consortium}). Funding for the DPAC has been provided by national institutions, in particular the institutions
participating in the {\it Gaia} Multilateral Agreement. 
\textbf{Funding:} T.R. and A.I.S. have been funded by the European Research Council (ERC) under the European Union's Horizon 2020 research and innovation programme (grant agreement No. 715947). S.K.S. acknowledges support by the BK21 plus programme through the National Research Foundation (NRF) funded by the Ministry of Education of Korea. E.M.A.G. acknowledges support by the International Max-Planck Research School (IMPRS) for Solar System Science at the University of G\"ottingen. 
\textbf{Author contributions:}  T.R., A.I.S., and S.K.S. conceived the study. A.I.S. and S.K.S. supervised the project. T.R. analyzed the Kepler data. B.T.M. investigated instrumental effects, and performed cross-matching the Kepler and Gaia catalogs. A.I.S., S.K.S., N.A.K., R.H.C, and E.M.A.G. contributed to the analysis of the data. T.R., A.I.S., S.K.S., and B.T.M. wrote the paper. All authors reviewed the manuscript.
\textbf{Competing interests:} There are no competing interests to declare.
\textbf{Data and materials availability:} The PDC-MAP Kepler data used in this study can be downloaded at \url{https://edmond.mpdl.mpg.de/imeji/collection/1qSQkt89EYqXAA2S}. Kepler data reduced with the PDC-msMAP pipeline are available at the Mikulski Archive For Space Telescopes, \url{https://archive.stsci.edu/pub/kepler/lightcurves/}. The sunpsot data was taken from \url{https://solarscience.msfc.nasa.gov/greenwch/sunspot\_area.txt}. The SATIRE-T2 data can be found at \url{http://www2.mps.mpg.de/projects/sun-climate/data/SATIRE-T2_TSI.txt} and the VIRGO level~2 1-minute data was taken from \url{ftp://ftp.pmodwrc.ch/pub/data/irradiance/virgo/1-minute_Data/}.

\subsection*{Supplementary Materials}
Material and Methods \\
Figs.~S1 to S10 \\
References (34-61) \\
Data~S1 (machine-readable data tables and software scripts)

\newpage
\includepdf{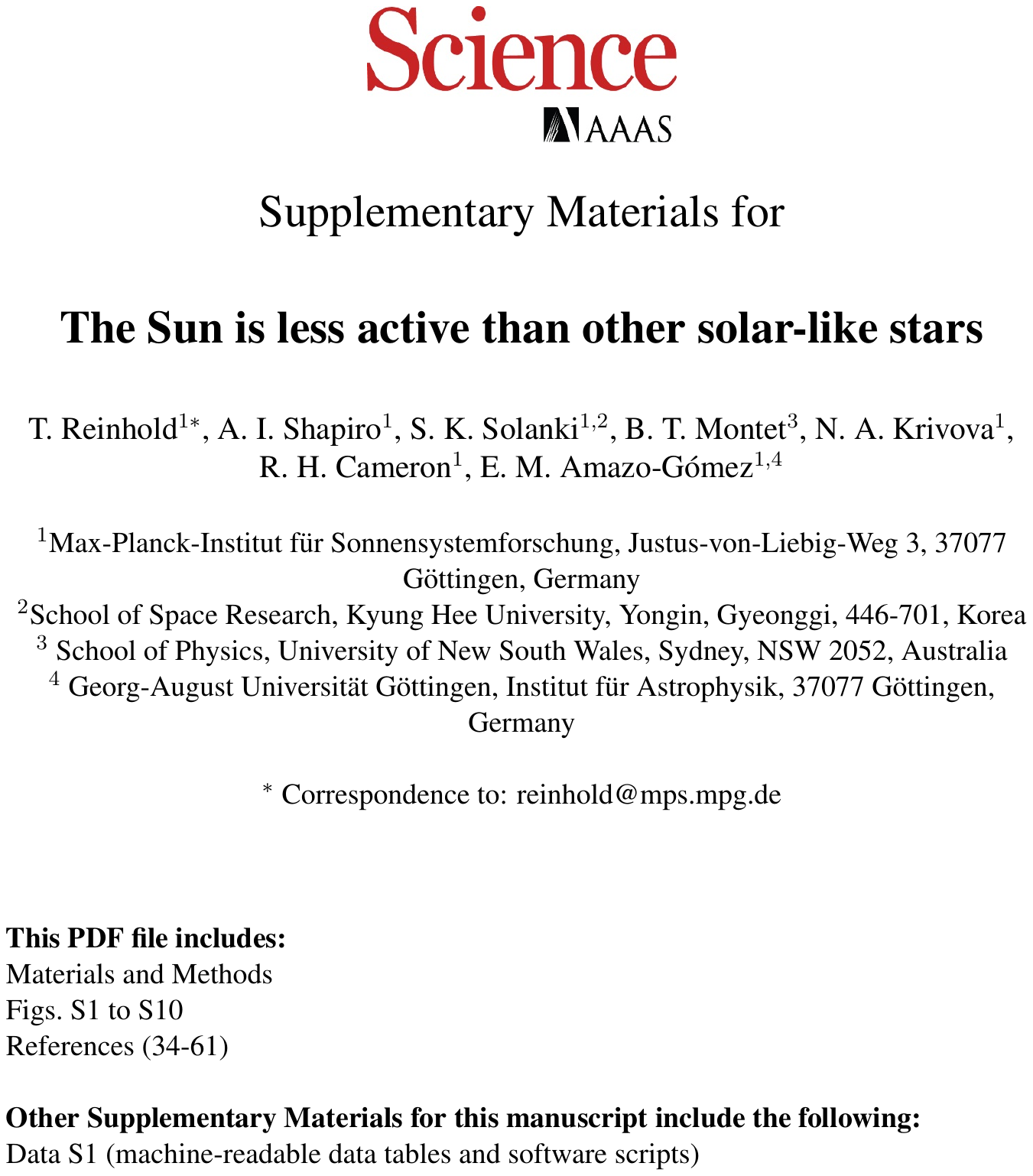} 

\newpage
\renewcommand{\thefigure}{S\arabic{figure}}
\setcounter{figure}{0}
\setcounter{page}{1}

\newcounter{defcounter}
\setcounter{defcounter}{0}
\newenvironment{myequation}{%
\addtocounter{equation}{-1}
\refstepcounter{defcounter}
\renewcommand\theequation{S\thedefcounter}
\begin{equation}}
{\end{equation}}

\noindent
\textbf{Materials and Methods}\\

\noindent
\underline{Isochrones} \\
Isochrones are evolutionary tracks of constant age along the HRD. We used PARSEC models \cite{Bressan2012,Chen2014,Chen2015,Tang2014,Marigo2017,Pastorelli2019,Weiler2018,Chen2019,Groenewegen2006,Kroupa2001,Kroupa2002} to fit the isochrones to the HRD in Fig.~\ref{HRD} (\url{http://stev.oapd.inaf.it/cgi-bin/cmd}). The position of the isochrone in the HRD depends on metallicity, i.e. the stellar brightness increases with metallicity. To restrict the samples to solar-like stars we selected stars that are confined between a lower isochrone of 4\,Gyr and metallicity of $[Fe/H]=-0.8$~dex, and an upper isochrone of 5\,Gyr and metallicity of $[Fe/H]=0.3$~dex. The chosen metallicity range is consistent with the catalog metallicities \cite{Mathur2017}. \\

\noindent
\underline{Kepler data processing} \\
Kepler data are released in "quarters" spanning $\sim$90~days. For each quarter the data were processed through a series of pipelines which can produce systematic effects which could create artificial signals of the amplitude considered here. Kepler records data for each star in typically 20-30 pixel "postage stamp" regions for which photometry is stored at 30-minute cadence. In producing light curves, a subset of these pixels, typically 4-6 in number, are summed to create Simple Aperture Photometry (SAP) time series light curves. These regions are similar in size to the stellar point spread function (PSF), so any small changes in the telescope pointing or temperature which affect the PSF shape or position will affect the amount of light that falls in the aperture. Typically these signals are dominated by the apparent movement of stars through differential velocity aberration (DVA) \cite{Kinemuchi2012} which produces a smooth, consistent variation throughout an observing quarter.

These long-term trends can be mitigated. The effects of shared systematics across the detector can be removed by regression fitting of shared signals on nearby targets. This method is the basis for the Kepler Pre-Search Data Conditioning (PDC) pipeline. The pipeline, while removing DVA-induced long-term trends, also can remove astrophysical long-term trends that are similar in length to the observing quarter \cite{Stumpe2012,Smith2012}. 

\begin{figure}[t]
    \centering
    \includegraphics[width=0.8\textwidth]{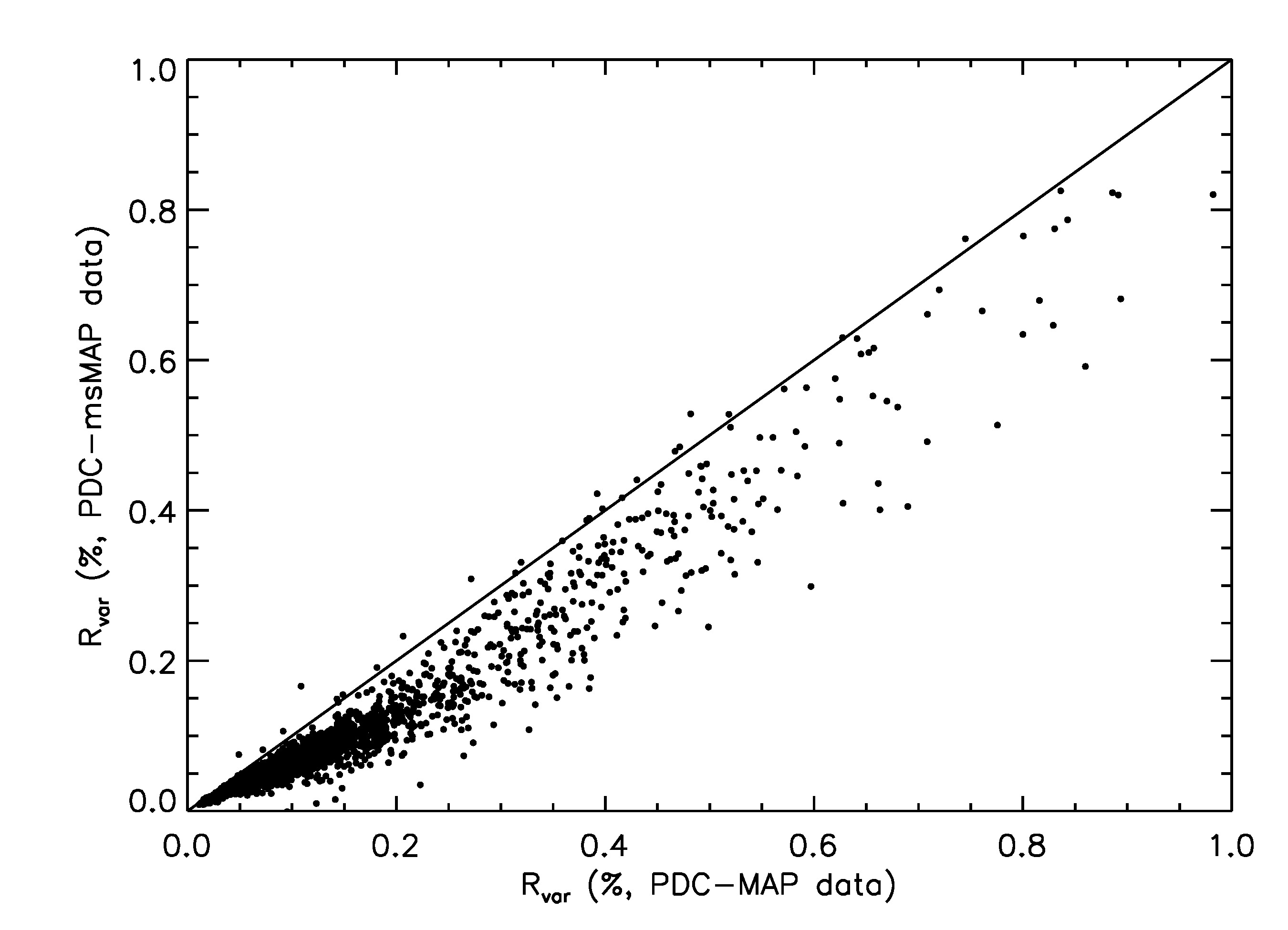}
    \caption{\textbf{Data comparison among different pipelines.} Comparison between $R_\text{var}$ values calculated using the PDC-MAP and PDC-msMAP pipelines. The black line indicates a 1:1 relationship.}
    \label{Rvar_old_new}
\end{figure}

\begin{figure}
  \centering
  \includegraphics[width=1.0\textwidth]{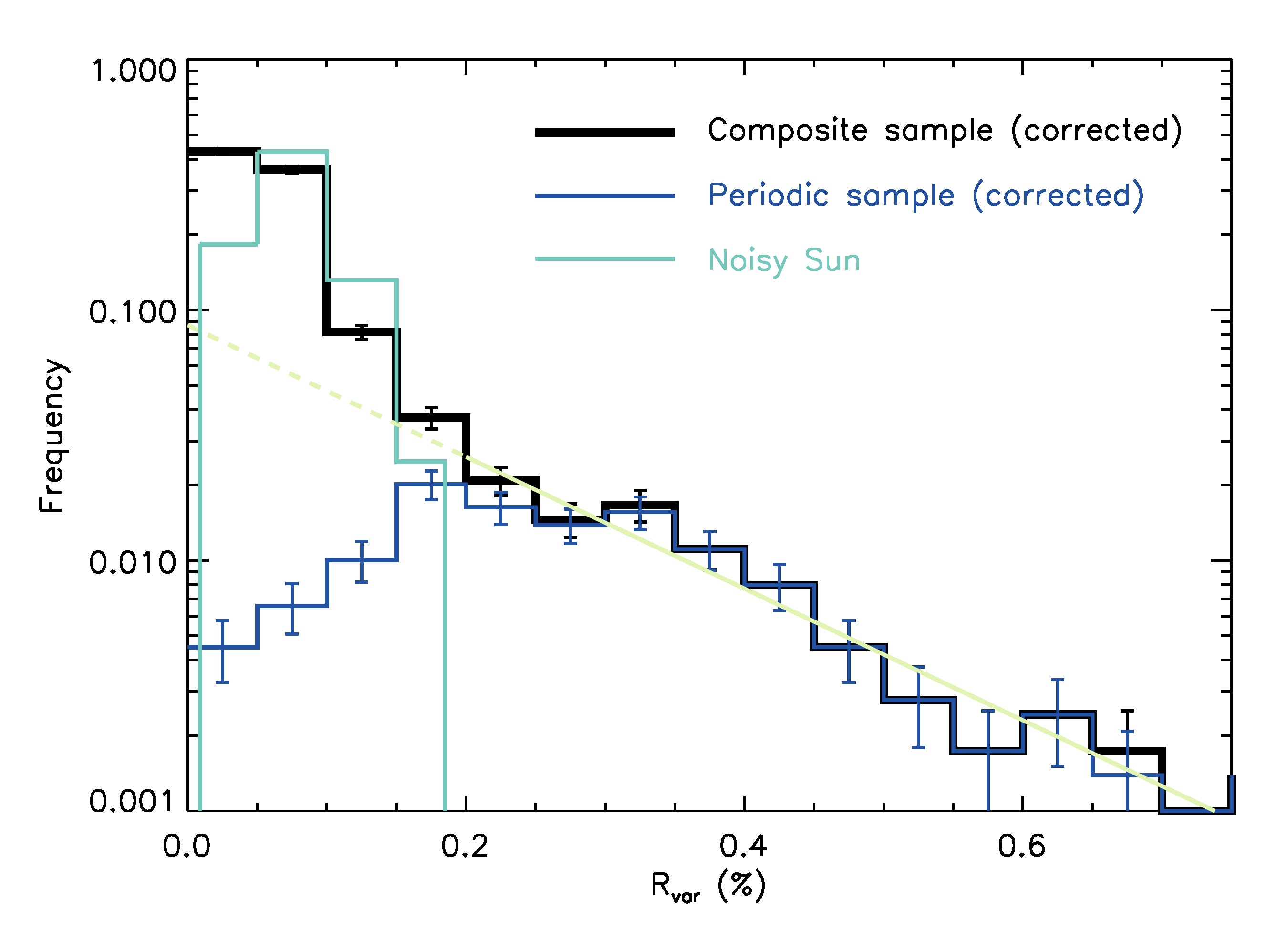}
  \caption{\textbf{Stellar variability distribution obtained using a different pipeline.} Same as Fig.~\ref{Rvar_dist} but using the  PDC-msMAP pipeline.}
  \label{Rvar_dist_new}
\end{figure}

We primarily used the PDC pipeline as produced by the Kepler team \cite{Smith2012}. We tested versions of the software from 8.0 to 9.2 in the analysis. The largest change between versions is that versions 9.0 and later use a PDC multi-scale Maximum a priori approach (PDC-msMAP), which applies a wavelet transform to separate the data into three channels which are independently treated for systematics \cite{Stumpe2014}. By separating the shared systematics by frequency, this pipeline reduces the risk of underfitting and performs better at separating astrophysical from instrumental signals. We find systematically larger $R_\text{var}$ values at the $\sim10-20$\% level when using version 8.3 of the pipeline than when using version 9.1. Fig.~\ref{Rvar_old_new} shows a comparison between the computed $R_\text{var}$ values for the two pipeline versions, and Fig.~\ref{Rvar_dist_new} shows the variability distribution using version 9.1 of the pipeline.

We compare the two pipelines and verify that the signals are not instrumental in nature by returning to the pixel-level data. As long-term systematics are driven by the movement of the star relative to the aperture, astrophysical signals should not be strongly dependent on the size of the photometric aperture while instrumental signals should be. For a test subsample of 13 randomly selected stars, each about twice as active as the Sun, we built a series of new photometric apertures and measured the brightness of the star over time for each aperture. While the long-term trends in the data vary with aperture, for each case we saw by eye a photometric signal which is not a function of aperture, and is therefore shared on all pixels across the image of the star on the detector. In all of these cases, the signals are consistent in amplitude with that recovered from the PDC pipeline, so must be astrophysical and associated with the star in question rather than instrumental.

\begin{figure}
  \centering
  \includegraphics[width=1.0\textwidth]{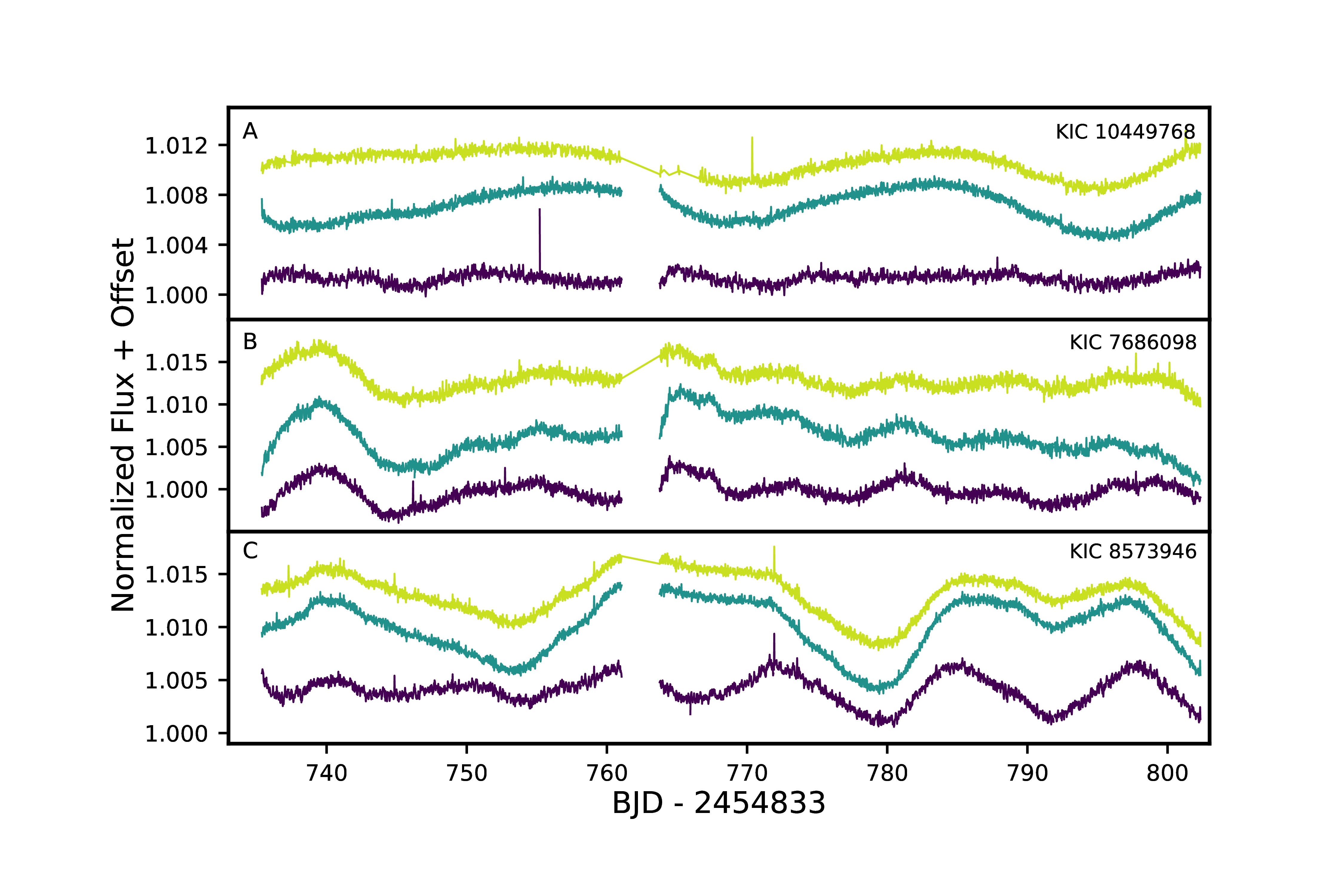}
  \caption{\textbf{The effect of different data reduction pipelines.} Sample light curves for three high variability stars, with KIC ID numbers listed in each panel. The photometry is taken from the PDC-msMAP data (purple curves), PDC-MAP data (dark green curves), and after our aperture selection and long-term systematics removal by searching for variability shared across different aperture choices (light green curves). In each case, the overall magnitude of the variability estimated in our pipeline is larger than from the PDC-msMAP pipeline but less than the PDC-MAP pipeline. In many cases, PDC-msMAP appears to remove true variability, while PDC-MAP tends to undercorrect instrumental long-term trends in the data.}
  \label{fig:lightcurvecomp}
\end{figure}

In comparing the light curves with different apertures, we searched for signals which are largely invariant to aperture size, unlike the effects of differential velocity aberration which tend to move a star into or out of its targeted aperture. We used this series of light curves to estimate the true astrophysical variability and compared it to the PDC-MAP and PDC-msMAP light curves. We find the PDC-msMAP light curves at times overcorrect stellar variability. However, the PDC-MAP light curves are often undercorrected and contain additional long-term trends over a month of observations. Examples of these are shown in Fig.~\ref{fig:lightcurvecomp}. These more often also contain variability induced by thermal variations as the spacecraft returns to thermal equilibrium after a data downlink. Therefore, in many cases the true level of astrophysical variability appears to be lower than presented in the PDC-MAP light curve but higher than in the PDC-msMAP light curves. \\

\noindent
\underline{Background stars} \\
To exclude possible contamination of the variability distribution by undetected background stars, which could be several times fainter than the main target but exhibit high intrinsic variability, we searched the Gaia DR2 catalog \cite{Gaia_DR2} for sources within a 4~arcsec radius around the location of each target in our sample. This radius corresponds to the scale of the pixels in the Kepler focal plane. Depending on the star, this search returns up to five different Gaia sources (including the target itself). We have selected stars brighter than 15th magnitude (in the Kepler band), while Gaia is limited to 21th magnitude, so we expect the Gaia catalog to be complete down to magnitudes much fainter than the main target. The median brightness of stars in our periodic sample is 14.5~mag, while the median brightness of nearby background stars is 18.6~mag.

To determine whether the high variability regime in the $R_\text{var}$ distribution is more affected by background stars, we compared the number of stars with more than one Gaia source for variabilities above and below $R_\text{var}=0.2\%$: we found that 10.3\% of the stars with $R_\text{var}>0.2\%$ and 11.4\% of the stars with $R_\text{var}<0.2\%$ have more than one background source. Because the above numbers are similar, we can exclude that the high variability tail results from unresolved, highly-variable backgrounds stars because this would have affected both samples.

Typical Kepler apertures used by the pipeline in the production of light curves for stars in our sample use data from 4-6 pixels, depending on the brightness of the star \cite{Haas2010}. These apertures change from observing quarter to quarter based on the orientation of the telescope and properties of the instrument channel. Our analysis allows us to directly exclude contamination from background stars in the inner part of the aperture for nearly 90\% of the foreground stars. Contamination of the stellar variability coming from the outer part of the aperture would be substantially larger in some quarters than others, leading to rapid, sudden changes in the observed variability on quarterly timescales and with yearly periodicity. We do not detect this effect in our $R_\text{var}$ data, which provides us with confidence that the observed variability is due to the sample stars. 

For one target in the test subsample, KIC\,9210546, there is another target of approximately equal brightness in the photometric aperture (KIC\,9210535), which could be the cause of the rotation signal. We modeled the PSF of both stars on the detector using the PSF photometry fitting tools in the \texttt{eleanor} software package \cite{Feinstein2019}, which enables us to separate the contributions of the two stars. We find that the 0.47\% variability is associated with the target star rather than the nearby companion. \\

\noindent
\underline{Uncertainties in the effective temperatures} \\
Most of the effective temperature values from photometry have uncertainties of about 150--200\,K \cite{Mathur2017}, which are too small to impact our analysis. The periodic stars make only 12.7\% of our sample and the number of highly variable stars (i.e. $R_\text{var}>0.4\%$) is about 5\%. Hence, it is possible that the color-based algorithm \cite{Mathur2017} failed for highly variable stars and erroneously attributed them to the 5500--6000\,K temperature bin, while in fact being much cooler and thus generally more variable.

\begin{figure}[t]
    \centering
    \includegraphics[width=0.8\textwidth]{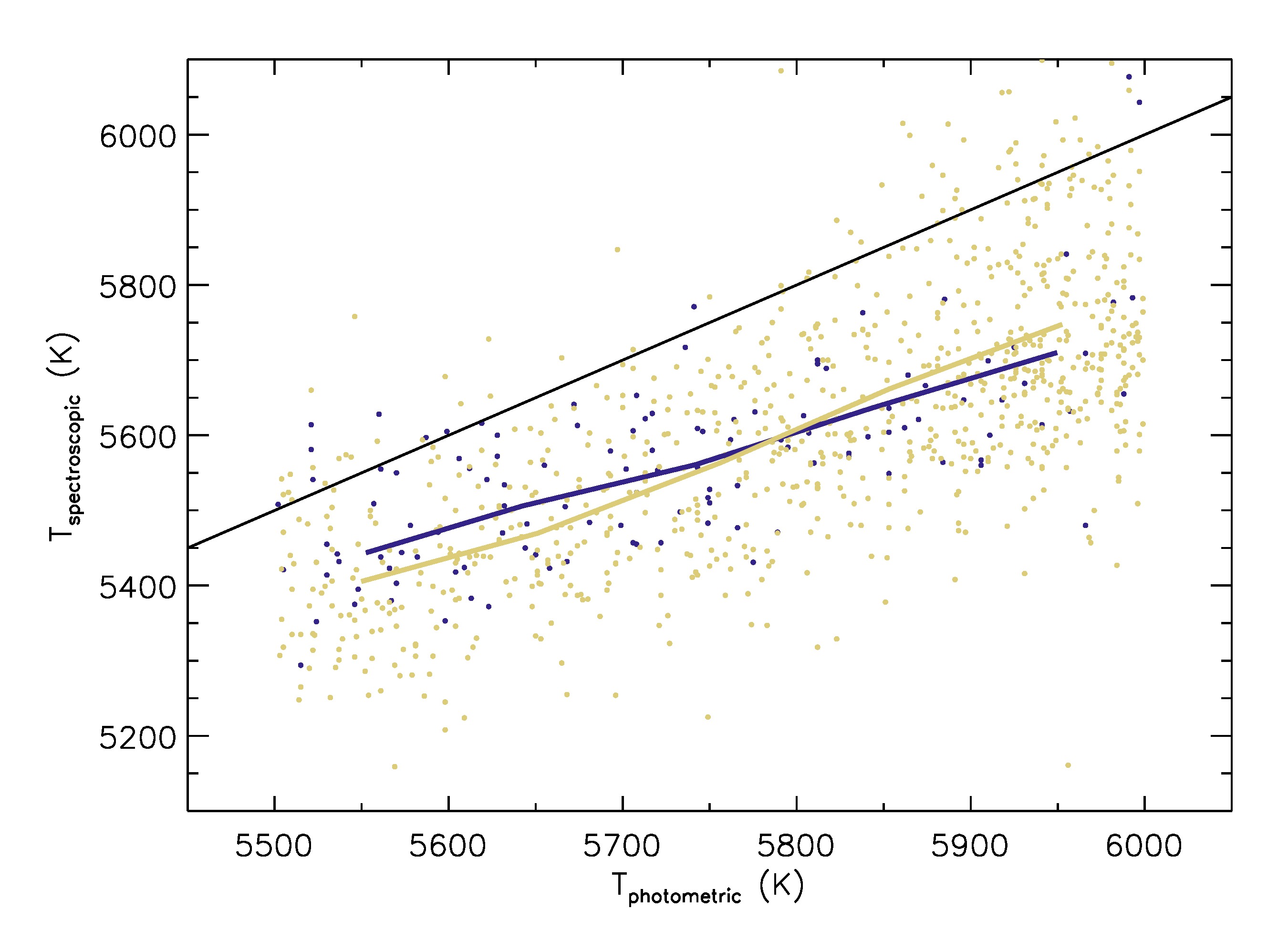}
    \caption{\textbf{Effective temperature differences between 
    photometric and spectroscopic catalogs.} Comparison between photometric effective temperatures \cite{Mathur2017} and 
    spectroscopic effective temperatures \cite{Zong2018}. Yellow and purple dots show the non-periodic and periodic stars, respectively, and the yellow and purple lines are the corresponding binned values. The black line indicates a 1:1 relationship.}
    \label{fig:T}
\end{figure}

To exclude this scenario we considered temperatures determined from low-resolution spectra, taken from the Large Sky Area Multi-Object Fiber Spectroscopic Telescope (LAMOST)-Kepler database \cite{Zong2018}. This contains LAMOST temperatures for 123 periodic stars from our sample and 836 non-periodic stars. The comparison between both sets of temperatures is shown in Fig.~\ref{fig:T} for the periodic and non-periodic stars. The LAMOST spectroscopic temperatures appear to be systematically lower (by roughly 200\,K) than those from photometry \cite{Mathur2017} for both the periodic and non-periodic samples. A similar offset has been found between the photometric temperatures \cite{Mathur2017} and those in the original Kepler Input catalog (KIC) \cite{Brown2011}. The photometric temperatures \cite{Huber2014,Mathur2017} were recalibrated to the infrared flux method (IRFM, \cite{Pinsonneault2012}) temperature scale using a revised color-temperature-metallicity relationship \cite{Casagrande2010}. The IRFM scale has been calibrated against solar twins and the Sun with an accuracy of a few tens of Kelvin [\cite{Casagrande2010}, their table~7]. Consequently, the photometric temperature scale \cite{Mathur2017} should be used for selecting stars with near-solar effective temperatures.

There is no systematic offset between the periodic and non-periodic stars (Fig.~\ref{fig:T}). The mean temperature offset between LAMOST and photometric \cite{Mathur2017} temperatures is 168\,K for the periodic and 189\,K for the non-periodic stars. Among the periodic stars, the offset does not show a specific dependence on the variability. For example, the mean offset for the 74 periodic stars with $R_\text{var}<0.4\%$ is 161\,K, while the mean offset for the 49 periodic stars with $R_\text{var}>0.4\%$ is 177\,K. We conclude that the LAMOST data indicate there is no systematic temperature shift between the periodic and non-periodic stars. \\

\noindent
\underline{Solar $R_\text{var}$ estimation} \\
Solar photometric variability has been shown to correlate well with other proxies of solar magnetic activity \cite{Salabert2017}. We quantify stellar photometric variability by the variability range $R_\text{var}$ \cite{Basri2010,Basri2011}. 
The Kepler light curves were rebinned from 30-minute to 3-hour cadences, and cleaned of outliers by discarding all data points deviating by more than six times the median absolute deviation. For a given star, the variability range was calculated for each quarter individually. To account for instrumental trends in certain quarters, we adopt the median of all quarters (numbered Q0--Q17) as a measure of variability over the total observing time span of 4~years.

\begin{figure}
  \centering
  \includegraphics[width=1.0\textwidth]{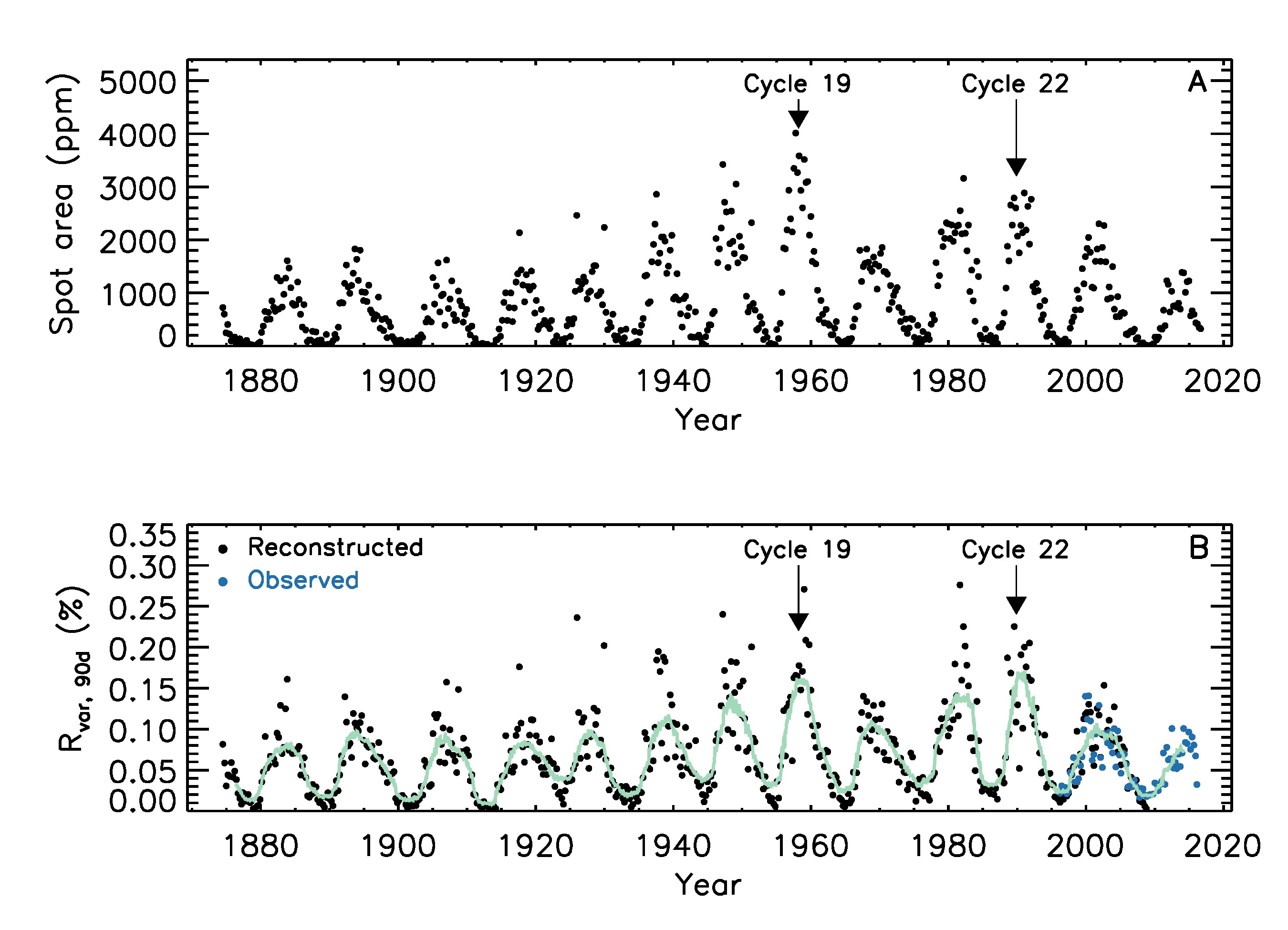}
  \caption{\textbf{140~years of solar activity data. (A)} Spot area coverage of the solar surface. \textbf{(B)} Variability range $R_\text{var, 90d}$ calculated over the 90-day intervals of the observed (blue) and reconstructed (black) total solar irradiance (TSI) time series. The solid green line shows the variability range $R_\text{var, 4yr}$ calculated as median of all $R_\text{var, 90d}$ values over the 4-year interval of the noise-free TSI time series (see Fig.~\ref{Rvar_area}A).}
  \label{Rvar_area_time_series}
\end{figure}

The photometric variability of the Sun was computed in a similar way. Fig.~\ref{Rvar_area_time_series}A shows consecutive 90-day averages of the solar surface sunspot area \cite{Hathaway2015_LR}, which is a direct measure of magnetic activity. Fig.~\ref{Rvar_area_time_series}B shows the variability range $R_\text{var, 90d}$ computed from contemporaneous 90-day segments of the full $\sim$140~years time series of the reconstructed TSI data from the Spectral And Total Irradiance REconstruction (SATIRE-T2) model \cite{SAT_T2_1}, and 20~years of the observed TSI data taken by the Variability of solar IRradiance and Gravity Oscillations (VIRGO) experiment \cite{Froehlich1995}. The variability range strongly correlates with the total sunspot coverage: both quantities are in phase and vary with the 11-year solar cycle. The reconstructed TSI data end in November 2008, while the measured TSI data are taken until March 2016. Fig.~\ref{Rvar_area_time_series}B shows the variability range $R_\text{var, 4yr}$, which is calculated exactly as the Kepler $R_\text{var}$ values: each 4-year segment is divided into sub-segments of 90~days, and the median of all individual $R_\text{var, 90d}$ values is taken as the $R_\text{var, 4yr}$ value.

TSI variability is not exactly equivalent to the solar brightness variability as it would be observed in the Kepler passband. The Sun is observed from near its equatorial plane (the angle between the solar equator and the ecliptic is $\sim 7.25^\circ$) while stars are observed at random inclinations (i.e. the angle between the line-of-sight direction of the observer and the stellar rotation axis). We employed the SATIRE-based model \cite{Nemec2020} to determine whether these two factors affect our analysis. We found that solar $R_\text{var}$ values in the Kepler passband are approximately 10\% larger than those obtained from TSI. Inclination-averaged solar $R_\text{var}$ values are 15\%-20\% smaller than those obtained from observations made from the solar equatorial plane. By simultaneously neglecting both, we overestimate solar variability by less than 10\%. Because the effect is small and because TSI represents an upper limit on the true solar variability in the Kepler passband, we used the original VIRGO TSI data and the TSI reconstruction, avoiding any correction for passband and inclination. 

\begin{figure}
  \centering
  \includegraphics[width=1.0\textwidth]{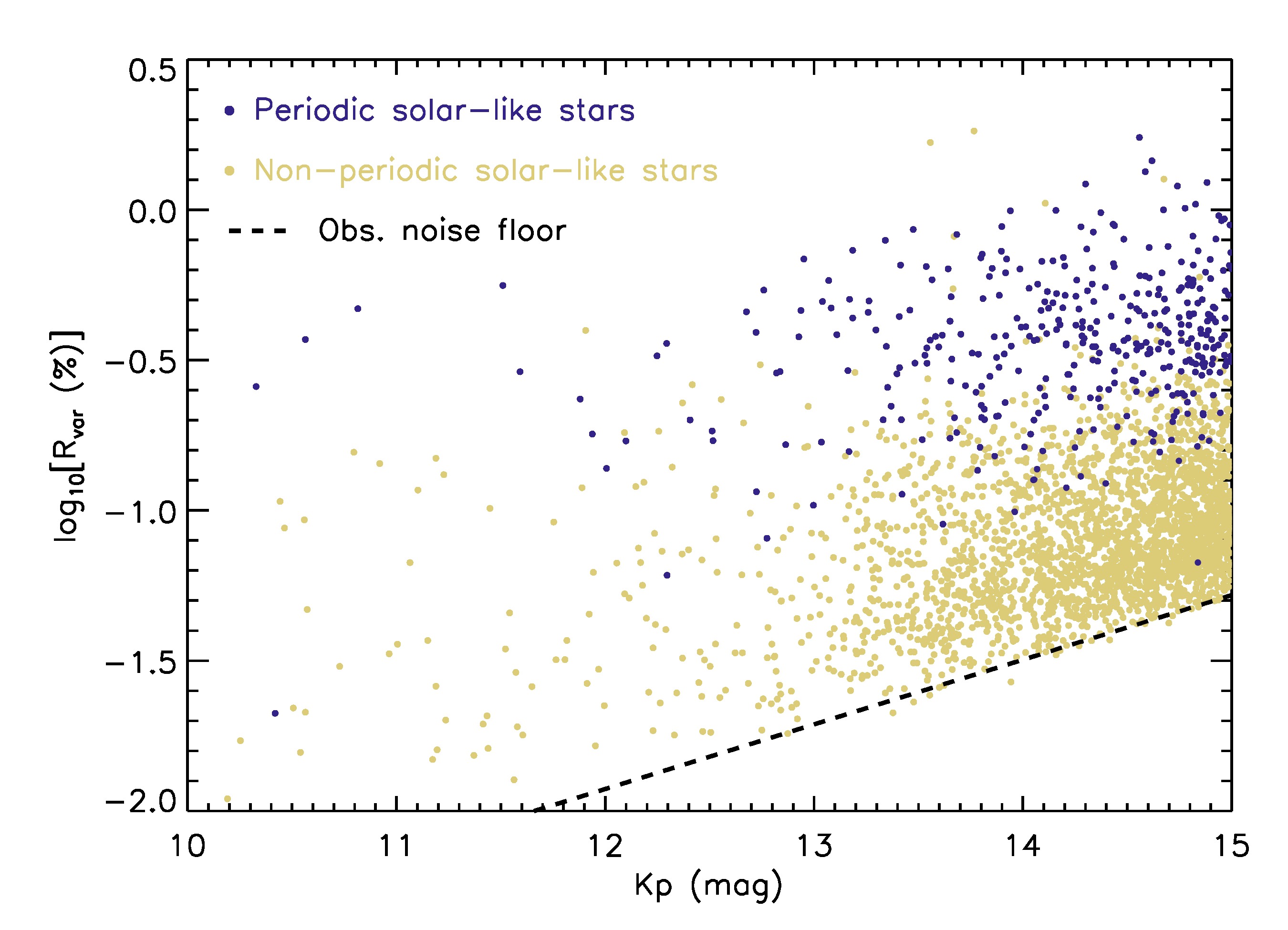}
  \caption{\textbf{Precision of the Kepler data.} Dependence of the variability range $\log_{10} R_\text{var}$ on apparent magnitude $Kp$ (in the Kepler band) of the periodic (purple) and the non-periodic (yellow) samples. The dashed black line marks our derived empirical noise floor.}
  \label{noise}
\end{figure}

To compare solar variability to that of the stars in our samples, we simulate how the Sun would appear if it was observed as a star of a given magnitude. The photometric precision achieved by the Kepler telescope strongly depends on the apparent magnitude of the observed stars. Fig.~\ref{noise} shows the dependence of the (logarithm of the) variability range $\log R_\text{var}$ on apparent magnitude in the Kepler band $(Kp)$ for the periodic and non-periodic samples. We fitted a linear model to the data, finding
\begin{myequation}
\log_{10} R_\text{var} = -4.507(\pm0.012) + 0.215(\pm0.001)\cdot Kp
\end{myequation}for the minimum values of $\log_{10} R_\text{var}$ for stars fainter than 14th magnitude. This indicates the photometric precision of Kepler, defining an empirical lower limit for the detectability of photometric brightness changes at a given magnitude.

To treat the Sun as a Kepler star, we ran a Monte-Carlo simulation selecting noise time series corresponding to random magnitudes between 10--15~mag. The random magnitudes used in the simulation are distributed in the same way as the sample magnitudes themselves, which contain many more faint stars (see Fig.~\ref{noise}). Consequently, many more runs are conducted with noise levels corresponding to fainter magnitudes. We then added a time series of Gaussian-distributed photon noise with standard deviation $\sigma$ to the TSI data. $\sigma$ was chosen to be appropriate to the magnitude picked by the Monte-Carlo simulation. The noise level $\sigma$ is linked to the computed $R_\text{var}$ values in Eq.~S1 as follows. The metric $R_\text{var}$ is computed by cutting the upper and lower 5\% of the sorted intensities and taking the difference between the maximum and minimum value of the middle 90\% of the data. Assuming Gaussian-distributed noise with zero mean and standard deviation $\sigma$, 90\% of all data points are contained within $\pm1.645\sigma$ around the mean value, so $R_\text{var}=3.29\sigma$.  

The noise floor given by Eq.~S1 is consistent with noise estimates for solar-like stars \cite{Gilliland2011}. Those authors used the standard Kepler noise metric, the combined differential photometric precision (CDPP) given in parts per million (ppm) instead of $R_\text{var}$ [\cite{Gilliland2011}, their figure~4]. Given a star with 14th magnitude, those authors binned the data to 6.5-hour cadences and estimated a noise $\log [CDPP (ppm)] \approx 1.7$.
Transferring this noise value to $R_\text{var}$ yields $R_\text{var}(3\,h, Kp=14\,mag) = 10^{1.7} \cdot \sqrt{6.5\,h/3\,h}\cdot 3.29 \approx 243\,ppm$.  Using Eq.~S1 we find $R_\text{var}(3\,h, Kp=14\,mag)=318\,ppm$. For a star with 15th magnitude, \cite{Gilliland2011} found $R_\text{var}(3\,h, Kp=15\,mag)=386\,ppm$, and Eq.~S1 yields $R_\text{var}(3\,h, Kp=15\,mag)=523\,ppm$.

\begin{figure}
  \centering
  \includegraphics[width=0.70\textwidth]{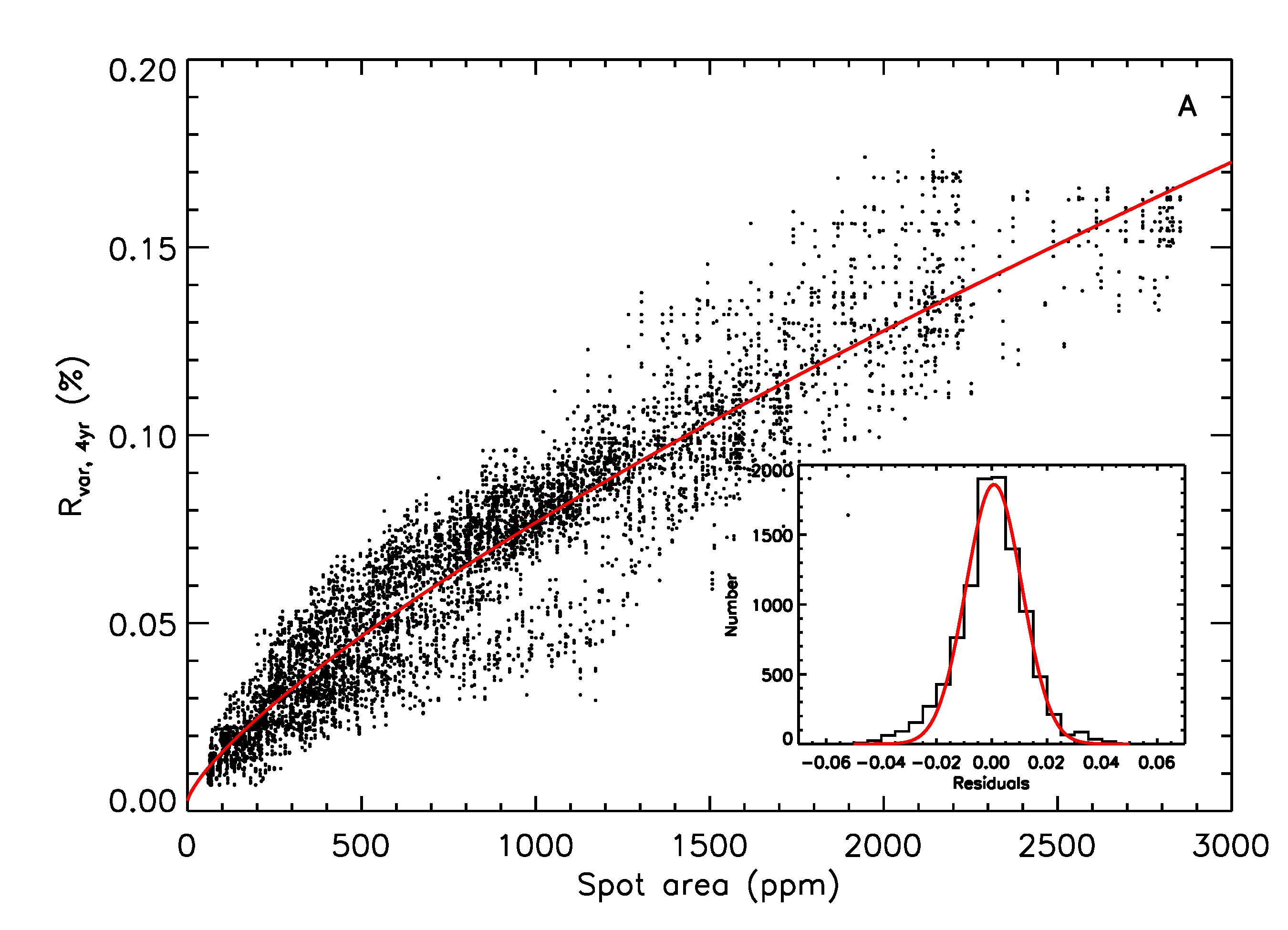}
  \includegraphics[width=0.70\textwidth]{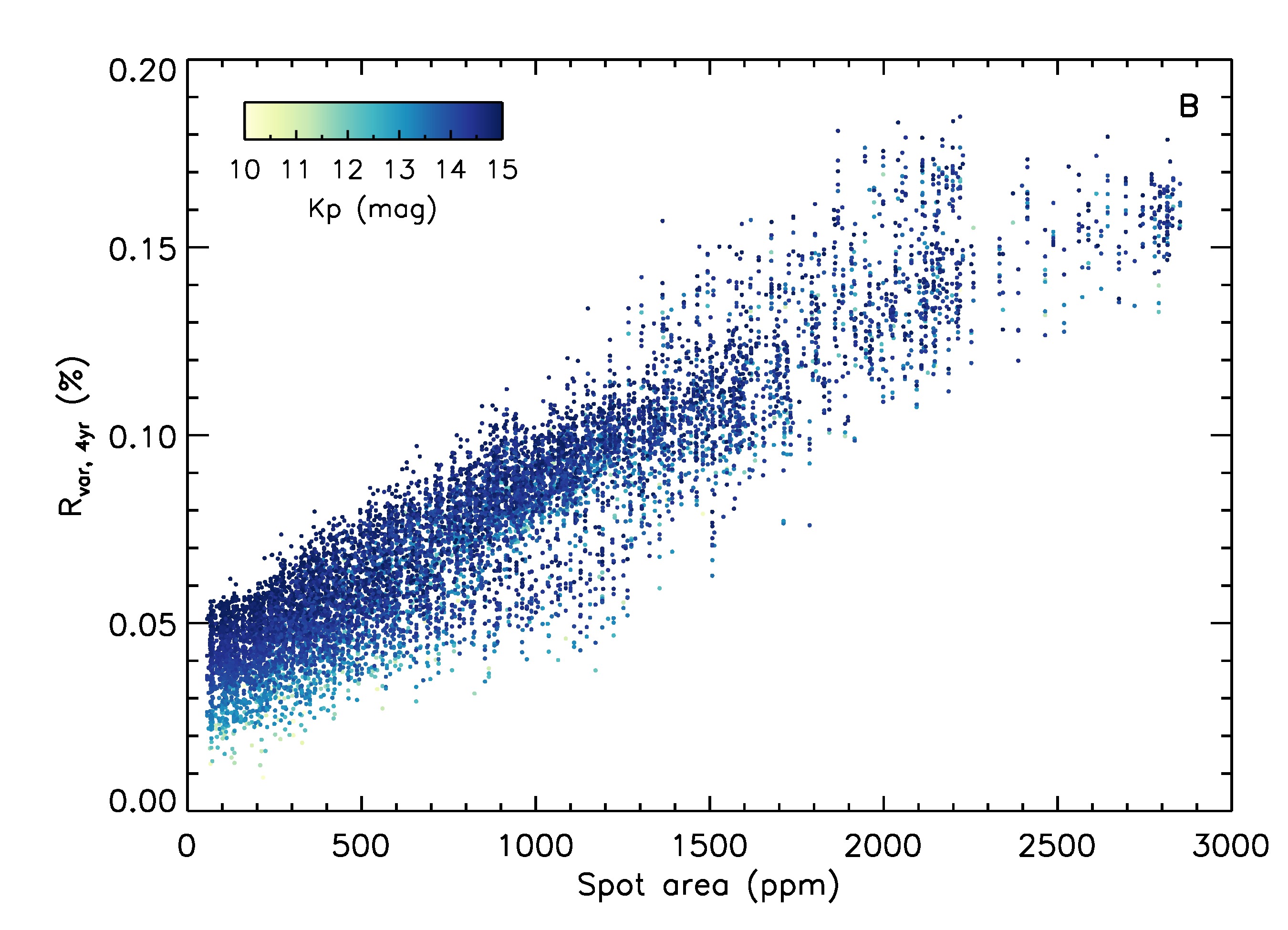}
  \caption{\textbf{Photometric vs. magnetic activity.} \textbf{(A)} Variability range $R_\text{var, 4yr}$ plotted against the solar spot area coverage of 10,000 randomly chosen 4-year segments of the noise-free TSI time series. The red curve shows a power law model $y=a_0+a_1\,x^{a_2}$ fitted to the data. The inset shows the histogram of the residuals (black), which are fitted with a Gaussian model (red). \textbf{(B)} Same quantities as in panel~\textbf{(A)} after adding magnitude-dependent noise to the 4-year segments of TSI data. The data are color-coded by the magnitudes used in the Monte-Carlo simulation. According to the magnitude distribution in Fig.~\ref{noise}, many more faint stars were considered. The distribution of these measurements of $R_\text{var, 4yr}$ is shown as the "Noisy Sun" in Fig.~\ref{Rvar_dist}.}
  \label{Rvar_area}
\end{figure}

We conducted 10,000 Monte-Carlo runs to compute the variability range $R_\text{var, 4yr}$ of the noisy Sun. Fig.~\ref{Rvar_area}A shows 
$ R_\text{var, 4yr}$ measurements for 10,000 randomly chosen 4-year segments of the noise-free TSI data against contemporaneous observations of sunspot areas. These are the same measurements which are shown in Fig.~\ref{Rvar_area_time_series}B. Fig.~\ref{Rvar_area} indicates the close connection between the magnetic activity and the brightness variability of a star. There is also a hysteresis pattern, similar to those found in previous analyses \cite{Salabert2017}, by comparing the photometric variability metric $S_\text{ph}$ \cite{Mathur2014} to different proxies of solar magnetic activity. The data can be fitted with a power law function $R_\text{var, 4yr} = 0.0024(\pm0.0007) + 0.00045(\pm0.00003)\cdot A_\text{s}^{0.74(\pm0.01)}$, where $A_\text{s}$ denotes the spot area. The distribution of the residuals, shown in Fig.~\ref{Rvar_area}A, is Gaussian. This shows that the scatter in $R_\text{var, 4yr}$ does not strongly depend on different levels of magnetic activity (as given by the spot area).

Fig.~\ref{Rvar_area}B shows the same dependence for the noisy Sun. The $R_\text{var, 4yr}$ values are offset by the magnitude-dependent noise $\sigma$. The noise increases with increasing magnitude, with greater affect on smaller values of $R_\text{var, 4yr}$. The (normalized) distribution of these noisy $R_\text{var, 4yr}$ measurements of the Sun is shown in Fig.~\ref{Rvar_dist} and Fig.~\ref{Rvar_dist_new}. \\

\begin{figure}
  \centering
  \includegraphics[width=0.7\textwidth]{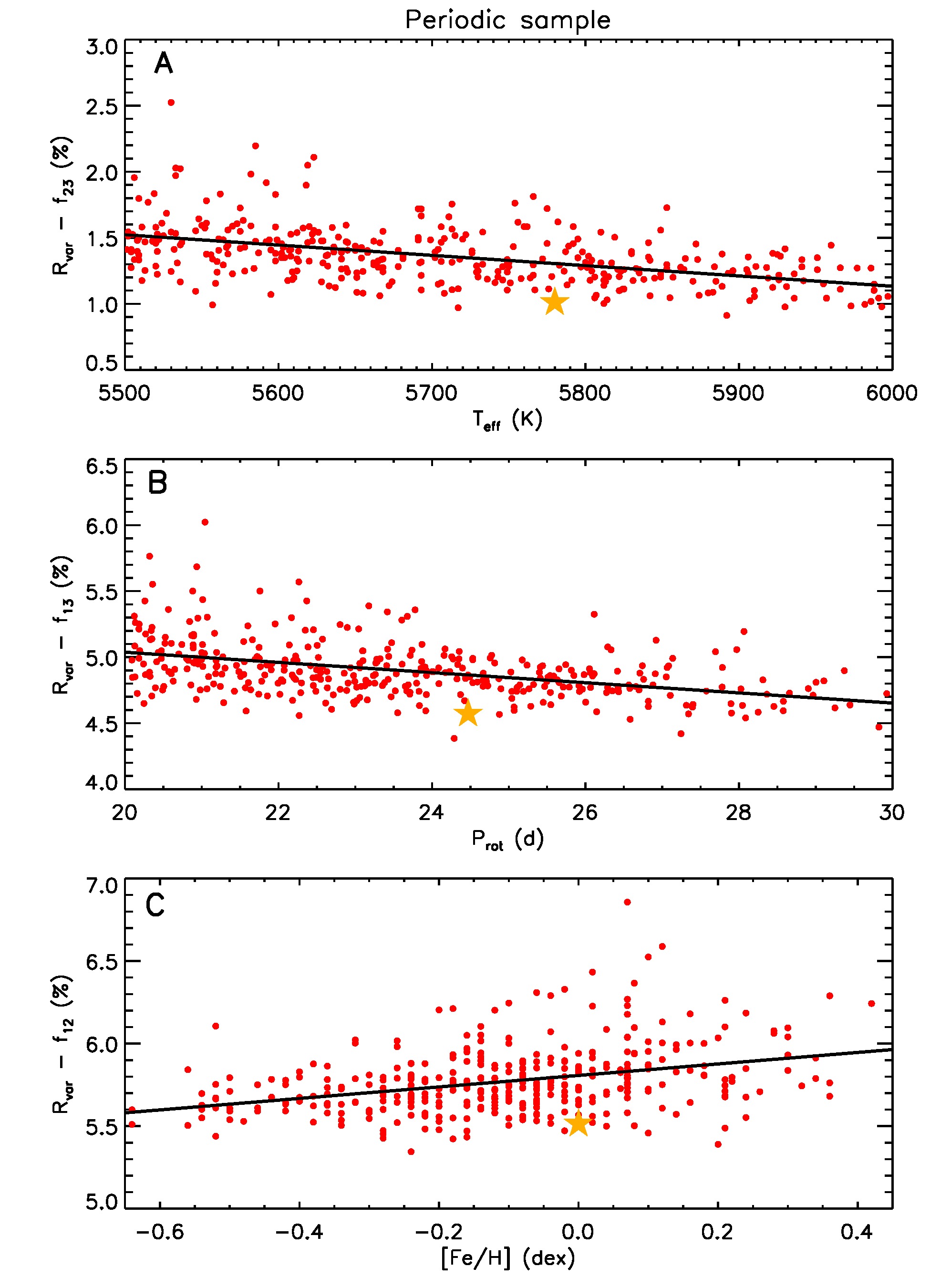}
  \caption{\textbf{Variability dependence on stellar fundamental parameters for the periodic sample.} Dependence of the variability range $R_\text{var}$ on \textbf{(A)} effective temperature $T_\text{eff}$, \textbf{(B)} rotation period $P_\text{rot}$, and \textbf{(C)} metallicity $[Fe/H]$ for the periodic sample. The data are fitted with a multivariate linear regression model $R_\text{var} (\%) = R_\text{var,0} + a_1\,(T_\text{eff}-T_{\text{eff},\odot}) + a_2\,(P_\text{rot}-P_{\text{rot},\odot}) + a_3\,([Fe/H]-[Fe/H]_\odot)$. The solid black line in each panel shows the model after subtracting the dependence of the other two parameters. E.g., the function $f_{23}$ in panel~\textbf{(A)} is defined as $f_{23}=a_2\,(P_\text{rot}-P_{\text{rot},\odot}) + a_3\,([Fe/H]-[Fe/H]_\odot)$, where the two function indices denote the model coefficients. The functions $f_{13}$ and $f_{12}$ are defined equivalently. The orange star indicates the Sun using its median variability $R_{\text{var},\odot}=0.07$\%.}
  \label{correction}
\end{figure}

\noindent
\underline{Dependence of $R_\text{var}$ on the fundamental parameters} \\
Fig.~\ref{correction} shows the variability range as a function of 
$ T_\text{eff}, P_\text{rot}, [Fe/H]$ for the periodic sample. We find that $R_\text{var}$ increases with decreasing effective temperature and rotation period (Fig.~\ref{correction}A-B), with a tendency to increase further towards lower temperatures and shorter rotation periods. $R_\text{var}$ increases with metallicity (Fig.~\ref{correction}C). The increase of $R_\text{var}$ with decreasing effective temperature and rotation period is consistent with the known relation between magnetic activity and the Rossby number, i.e., the ratio between the rotation period and the convective turnover time. The Rossby number decreases for faster rotation rates and smaller values of effective temperatures, and consequently the magnetic activity increases \cite{Noyes1984}. The observed trend with increasing metallicity is harder to explain. It might be connected with the effect of metallicity on the brightness of stellar magnetic features \cite{Witzke2018}, or might depend on the action of the stellar dynamo \cite{Karoff2018}. We fitted the data with a multivariate linear regression 
\begin{myequation}
    R_\text{var} (\%) = R_{var,0} + a_1\,(T_\text{eff}-T_{\text{eff},\odot}) + a_2\,(P_\text{rot}-P_{\text{rot},\odot}) + a_3\,([Fe/H]-[Fe/H]_\odot),
\end{myequation}where the coefficients are given by $R_\text{var,0} = 5.7981\pm0.3422$, $a_1=-0.0008\pm0.0001$, $a_2=-0.0383\pm0.0029$, and $a_3=0.3471\pm0.0331$ and the solar values are $T_{\text{eff},\odot}=5780$\,K, $P_{\text{rot},\odot}=24.47$\,d, and $[Fe/H]_\odot=0$. Although the dependencies on the fundamental parameters are rather weak, they might distort the variability distribution. Thus, the multivariate linear model is subtracted from the measured variabilities, and the variability range is referred to as the "corrected" $R_\text{var}$ in Fig.~\ref{Rvar_dist}.

\begin{figure}
  \centering
  \includegraphics[width=0.8\textwidth]{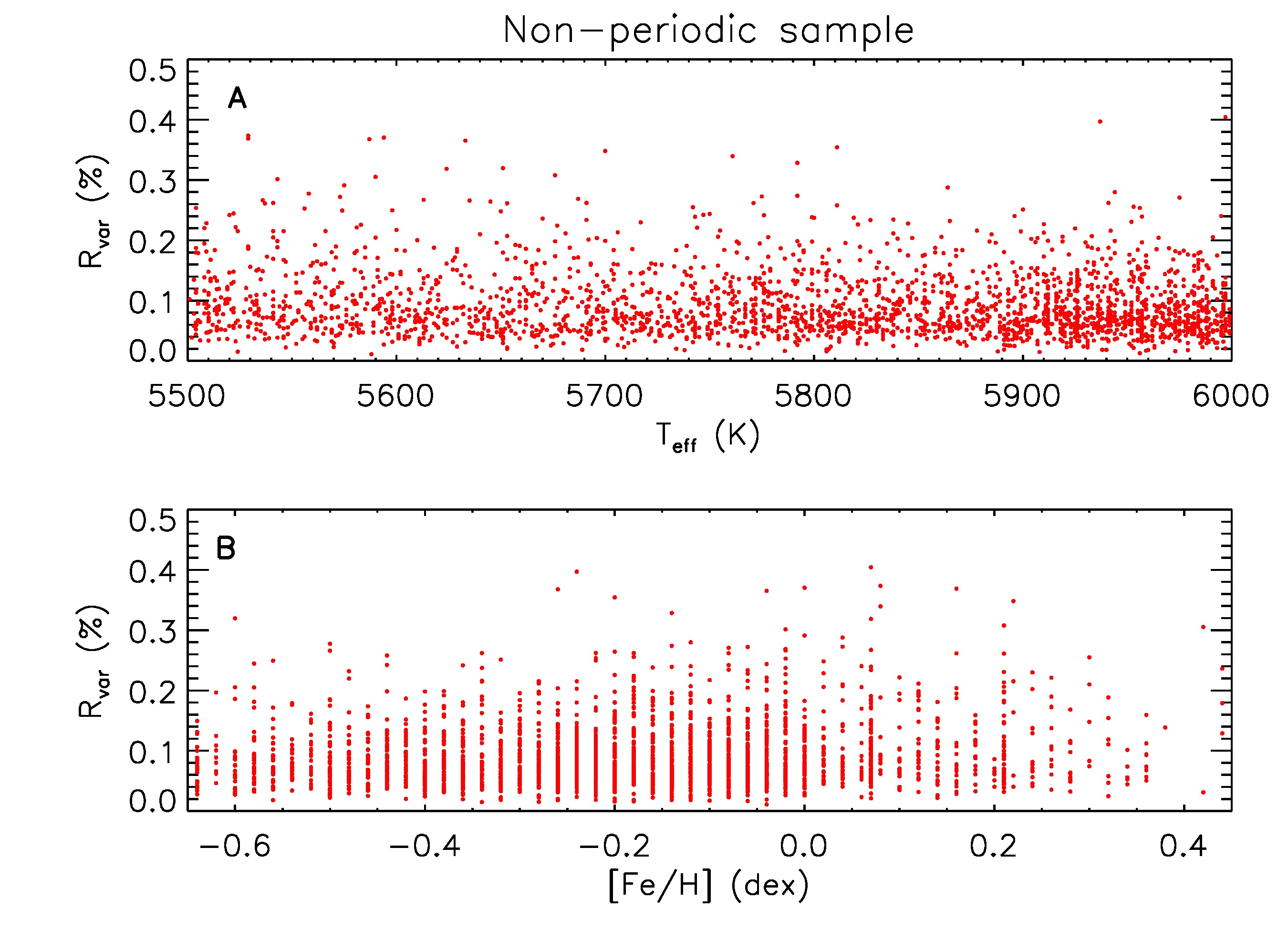}
  \caption{\textbf{Variability dependence on stellar fundamental parameters for the non-periodic sample.} Dependence of the variability range $R_\text{var}$ on \textbf{(A)} effective temperature $T_\text{eff}$ and \textbf{(B)} metallicity $[Fe/H]$ for the non-periodic sample.}
  \label{no_correction}
\end{figure}

In contrast to the periodic sample, the variability range of stars with unknown rotation period does not show a dependence on effective temperature or metallicity (Fig.~\ref{no_correction}). This might be attributed to, e.g., the absence of active regions on their surfaces, a geometrical effect (e.g. solar $R_\text{var}$ values decrease when the observer moves out of the solar equatorial plane \cite{Shapiro2016}), or a lower signal-to-noise ratio masking the dependencies seen for the periodic sample. A multivariate analysis was not carried out in this case, due to the lack of knowledge on the rotation period, whose substantial influence cannot be removed.

\begin{figure}
  \centering
  \includegraphics[width=1.0\textwidth]{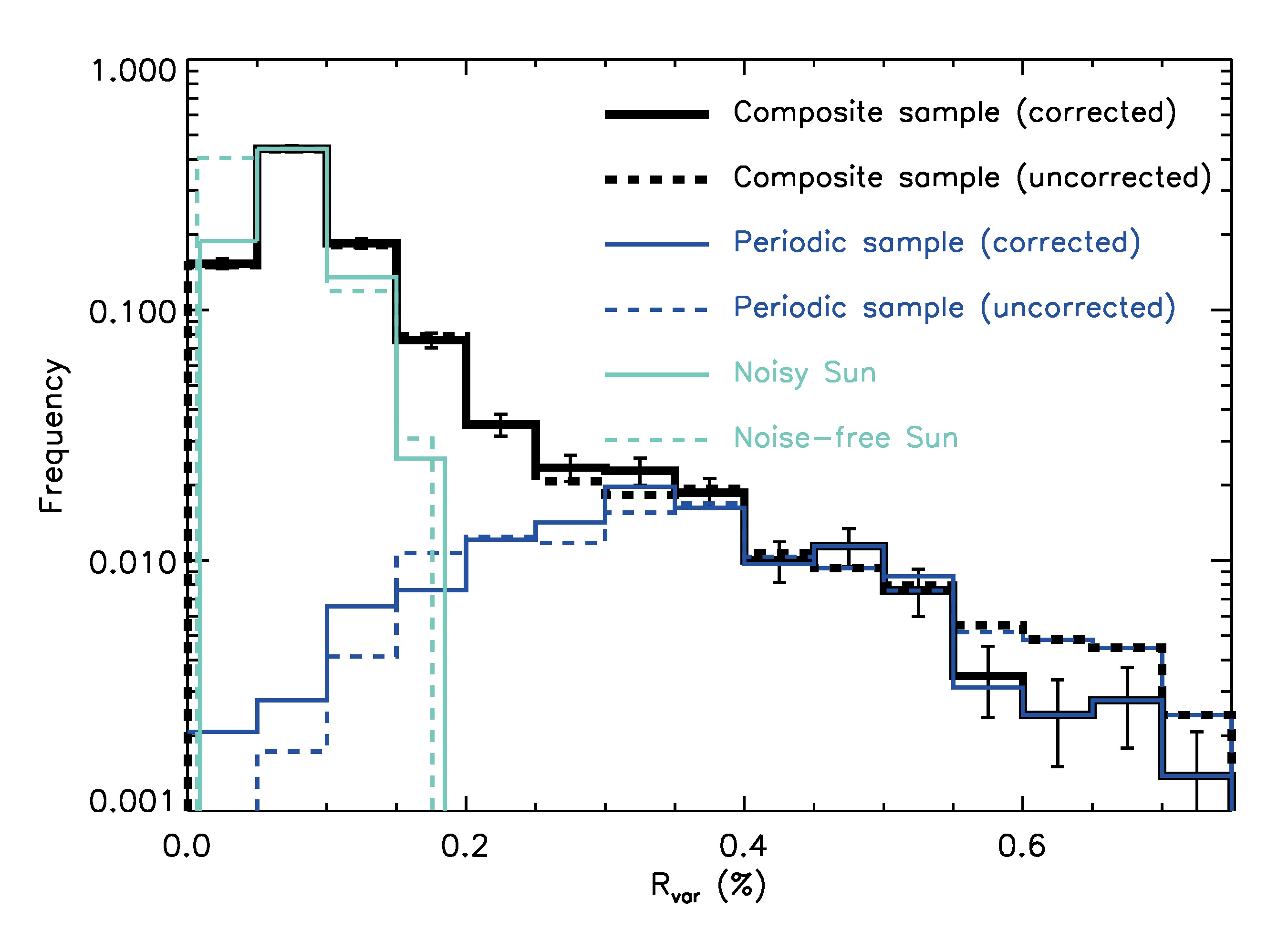}
  \caption{\textbf{Corrected vs. uncorrected variability.}
  Same as Figure~\ref{Rvar_dist}, but showing the distribution of $R_\text{var}$ for the corrected (solid) and uncorrected (dashed) samples, and the noisy (solid green) and noise-free (dashed green) Sun.}
  \label{Rvar_dist_corr}
\end{figure}

Fig.~\ref{Rvar_dist_corr} shows the impact of this correction on the variability distributions of the corrected and uncorrected $R_\text{var}$ values. As mentioned above, the dependencies of $R_\text{var}$ on the fundamental parameters are weak for the periodic stars, and minor for the non-periodic stars. Hence, the impact of the correction on the composite sample is rather small. Adding noise to the TSI data shifts the distribution of $R_\text{var}$ to higher values.

\newpage

\end{document}